\documentclass[aps,preprint,aps,showpacs]{revtex4}

\usepackage{epsfig}
\usepackage{graphicx}
\usepackage{amsmath}
\usepackage{amsfonts}
\usepackage{amssymb}%
\begin{document}
\title{Spin polarization tuning in Mn$_{x}$Fe$_{1-x}$Ge$_{3}$}
\author{A. Stroppa}
\affiliation{Faculty of Physics, University of Vienna, and Center
for Computational Materials Science, Universit\"at Wien, Sensengasse
8/12, A-1090 Wien, Austria}
\author{A. Continenza}
\affiliation{CNISM- Dipartimento di Fisica Universit\`a degli Studi
dell'Aquila, Via Vetoio 10 L'Aquila, Italy;}
\author{G. Kresse}
\affiliation{Faculty of Physics, University of Vienna, and Center
for Computational Materials Science, Universit\"at Wien, Sensengasse
8/12, A-1090 Wien, Austria}

\begin{abstract}

Experimentally, the intermetallic compound Mn$_{4}$FeGe$_{3}$ has
been recently shown to exhibit enhanced magnetic properties and spin
polarization compared to the Mn$_{5}$Ge$_{3}$ parent compound. The
present {\em ab-initio} study focusses on the effect of Fe
substitution on the electronic and magnetic properties of the
compound. Our calculations reveal that 
the changes on the
Fermi surface of the doped compound are remarkable and provide
explanations for the enhanced spin-polarization observed. Finally,
we show that it is indeed possible to tune the degree of
spin-polarization  upon Fe doping, thus making the
Mn$_{1-x}$Fe$_{x}$Ge$_{3}$ intermetallic alloy very promising for
future spintronic applications.
\end{abstract}

\maketitle

The ferromagnetic intermetallic Mn$_{5}$Ge$_{3}$ compound is a very
promising spin injector: it can be easily grown epitaxially on the
Ge(111) substrate\cite{epitaxy,weitering1,weitering2}  and can
provide  spin-polarization of about 42 \% while preserving
ferromagnetic ordering up to a Curie temperature (T$_{C}$) as high
as 296 K. Very recently, T. Y. Chen \emph{et al.} have shown that
replacing one Mn atom/f.u. with Fe may lead to a remarkable effect
on the spin-injection properties.\cite{mn4feg3} This immediately
opened the possibility for
 \emph{Spin Polarization  Engineering} (SPE) of
Mn$_{5}$Ge$_{3}$ by means of Fe doping. The reasons for the
remarkable spin polarization enhancement in Mn$_{4}$FeGe$_{3}$ are
not clear yet and await theoretical explanations.

Density Functional  calculations were performed using the VASP
package  within the  Generalized Gradient Approximation
(PBE-GGA).\cite{PBE} PAW  pseudopotentials\cite{PAW,PAWKresse} were
used for both Ge and TM-atoms: the semicore $3p$ states are
considered as valence (core) for Mn (Fe); the $3d$ states are frozen
in the core for Ge.  The kinetic energy cutoff used for the wave
functions was fixed to 350 eV. (4,4,6) $\Gamma$-centered $k$-points
were used for the self-consistent cycle, while a (12,12,14) k-point
grid was used for the calculation of the Fermi velocity. All the
atomic internal positions as well as  the volume and shape of the
unit cell were relaxed minimizing the {\it ab-initio} stress and
forces.

The Mn$_{4}$FeGe$_{3}$ compound can be represented as a solid
solution of Fe in Mn$_{5}$Ge$_{3}$, \emph{i.e.}
Mn$_{5-x}$Fe$_{x}$Ge$_{3}$ with $x=1$. For $x=0$, at ambient
conditions, Mn$_{5}$Ge$_{3}$ crystallizes in the hexagonal D8$_{8}$
type (space group P6$_{3}$/\emph{mcm}) with a unit cell containing
16 atoms: 10 Mn atoms in two inequivalent sites identified using the
Wyckoff notation [4M$_{I}$ in $4(d)$; 6M$_{II}$ in
  $6(g)$]  and 6
Ge atoms in $6(g)$ sites. The description of  the crystal structure
has been given elsewhere.\cite{mnge1} For $x=5$, Fe$_{5}$Ge$_{3}$ is
also hexagonal but belongs to the D8$_{2}$ symmetry type (space
group P6$_{3}$/\emph{mmc}), so that  the mutual solubility of the
end members could be  limited by their different crystal structure.
 The limit of solubility   appears to occur at $x$=1,
  or slightly beyond,\cite{exp1}
 resulting in the Mn$_{4}$FeGe$_{3}$ compound  that still preserves the
 same crystal structure as Mn$_{5}$Ge$_{3}$. Therefore,
Mn$_{4}$FeGe$_{3}$  can be described using
  an ordered supercell containing 16 atoms, \emph{i.e.}  8 Mn, 2 Fe and 6
Ge atoms.

We start our study considering one  Fe atom at both the M$_{I}$ or
M$_{II}$ site, and we found that Fe prefers to  occupy the M$_{I}$
site with an energy gain of 280 meV/cell with respect to the
M$_{II}$ site:
 thus,  in agrement with experiments,\cite{exp1,exp2,exp3}
the smaller and less electropositive iron atom substitutes
preferentially Mn on the $4(d)$ sites.\cite{webelements} This not
unexpected since occupation of the M$_I$ site allows the metallic
atoms to be closer\cite{exp4}, consistently with the Fe atoms having
smaller atomic radius than Mn. The $4(d)$ atomic positions (in
internal coordinates) lie in two different planes along the
hexagonal $c$-axis:
 (1/3,2/3,1/2),
(2/3,1/3,1/2), (1/3,2/3,0), (2/3,1/3,0) and they can be occupied by
two iron atoms in three different and non-equivalent configurations.
 We find
that,
in the lowest-energy state, the Fe atoms are located  on the sites
belonging to the $z=0$ or $z=c/2$ symmetry equivalent planes.

 \begin{table}
 \caption{Structural data (lattice constant $a$ and $c/a$ ratio, magnetic
 moments and heat of
 formation) for the structures considered compared with experiments where
 available.}
\label{tab1}
\begin{ruledtabular}
\begin{tabular}{lcccccccc}
 System        & $a$(\AA) &     $c/a$   & $\mu_{M_{I}}$ ($\mu_{B}$) & $\mu_{M_{II}}$ ($\mu_{B}$) & $\mu_{Ge}$
 ($\mu_{B}$)& $\mu_{T}$ ($\mu_{B}$/f.u.) &$\Delta H_{f} (eV/atom)$ \\

 Mn$_{5}$Ge$_{3}$   & 7.142         & 0.697  & 2.22          &  3.11  &  $-$0.16  & 2.70  & $-$0.140  & \\
 Exp\cite{exp4}     & 7.184              & 0.703  & 1.96          &  3.23&
    & 2.60 & &\\
    \hline
 Mn$_{4}$FeGe$_{3}$ & 7.134         & 0.690  & 1.75 (Fe), 2.35 (Mn)   &  3.05  &  $-$0.15  & 2.59  & $-$0.145 &  \\
 Exp\cite{mn4feg3} & 7.184         & 0.696  &                        &        &         & 2.35  &
 & \\
 Exp\cite{exp3}          & 7.138         & 0.702  &   1.55        &  2.45  &         & 2.10  &        &   \\
  \hline
 Fe$_{5}$Ge$_{3}$   & 6.967         & 0.692  & 1.58          &  2.17  &  $-$0.11  & 1.85  & $-$0.061 &  \\
\end{tabular}
\end{ruledtabular}
\end{table}
 In table~\ref{tab1}, we show the calculated lattice constants and magnetic moments
  of  Mn$_{5}$Ge$_{3}$, Mn$_{4}$FeGe$_{3}$ and
 the hypothetical Fe$_{5}$Ge$_{3}$ compound in the D8$_{8}$  crystal
 structure. The formation energy ($\Delta H_{f}$) is defined
  with respect to reservoirs of atoms in the pure bulk phases
  $\gamma$-Mn AFM1, FM bcc-Fe and diamond Ge; we have chosen the sign in such a way
   that stable compounds have
  negative   formation
 energies. For Mn$_{5}$Ge$_{3}$, the theoretical lattice
 constants $a$ and $c$ are  slightly underestimated with respect
 to the experimental values (0.5 \% and 1.4 \%, respectively).
 The calculated magnetic
moments  correctly reproduce the relative magnitude of the M$_{I}$
and M$_{II}$ moments, with the former carrying   smaller moments, in
agreement with experiments and previous theoretical
calculations.\cite{exp4,mnge1,mnge2}
 A small induced
 antiferromagnetic polarization is present in the cell, mostly localized
 on  Ge sites.\cite{exp4,mnge1,mnge2} This is true for all three cases considered.
 Upon  doping,
 the volume of the unit cell decreases by 1.4 \%;  the $c$-axis shrinks faster than the
 $a$-axis, thus decreasing the $c/a$ ratio. The M$_{I}$-M$_{I}$
 separation along the $z$-axis (\emph{i.e.} $c$/2) affects the length of the $c$-axis
  most: the smaller atomic size of Fe
 compared to Mn (the Fe and Mn covalent radii are 1.25 and 1.39
 \AA, respectively\cite{webelements}) is clearly responsible for
  the  contraction of the
 $c$-axis
upon Fe substitution. Our calculated lattice constants
  agree well with experimental values.\cite{mn4feg3,exp3}
 The magnetic moments at the M$_{I}$ sites are distributed as follows:
  Fe carries the lowest magnetic moment  (1.75 $\mu_{B}$), while
 Mn shows a slightly increased moment  compared to pure
 Mn$_{5}$Ge$_{3}$ (2.35 $\mu_{B}$). On the other hand, the M$_{II}$ magnetic
 moments are only marginally affected by Fe substitution.
  The experimental values
 shown in Tab.~\ref{tab1} are obviously average values for the
 two independent sublattices:
 the slight discrepancy between experimental and calculated values
could be ascribed to disorder effects which are
 not taken into account in the present calculation.

 It is interesting to explore the $x=5$ doping limit in the
 $D8_{8}$ phase.
 In  the Fe$_{5}$Ge$_{3}$ phase,  there is  a significant decrease of the length of
 both the $a-$ and $c-$axis by 2.4 and 3.1 \% respectively, while
   the c/a ratio is almost unchanged compared to the $x=0$ case.
   The iron magnetic moments decrease to 1.58 and 2.17 $\mu_{B}$ at
   M$_{I}$ and M$_{II}$ sublattice, leading to a sizable reduction of the
   total moment in the cell.
    Finally, concerning the  stability of the compounds,  we note that
     Mn$_{5}$Ge$_{3}$ and  Mn$_{4}$FeGe$_{3}$ are quite stable
     ($\Delta H_{f}$ are -0.140 and -0.145 eV/atom respectively), whereas
      Fe$_{5}$Ge$_{3}$ has only a slightly negative formation energy (-0.061 eV/atom).





We now focus on the electronic structure of the intermetallic
  compounds considered.
  According to Ref.~\onlinecite{exp2}, strong metal-germanium and metal-metal covalent bonding
  is  present in this phase.
  Broad, primarily $sp$  bonding  and  antibonding  bands form due to covalent
  interactions between Mn and Ge. Since there are three germanium atoms per formula unit, each contributing four
  orbitals with a spin degeneracy of two per orbital, these bands can accomodate 24 electrons.
  Assuming formal valences (Mn,Fe)$^{+3}$ at $4(d)$,
  (Mn,Fe)$^{+2}$ at $6(g)$, Ge$^{4-}$ at
  $6(g)$,\cite{exp2} the bonding band is filled with six electrons from the two $4(d)$ metal atoms,
   six from the three $6(g)$
  metal atoms, and twelve from the three metalloid atoms (in total 24). At the Fermi level,
    a large density of $d$-like states is present.
  Therefore, we   expect that there will be sheets at the
  Fermi surface with large and ''heavy"-like areas, made up predominantly of electrons
    relatively localized on partially occupied $d$-like states and, thus,
  with large effective masses and low  Fermi velocity.
 On the other hand, there will be also   small and "light"-like areas of the Fermi-surface sheets
  with $sp$ character, made up by highly mobile
 electrons with delocalized wave functions, small effective masses,
 and high Fermi velocity. Thus, it is expected that the
 $d$-likes state dominate the density of states at the Fermi level, N(E$_{f}$),
  while $sp$ electrons do contribute
 substantially to the average value of the Fermi velocity.

\begin{figure}
\hspace{0.truecm}\includegraphics[width=0.5\textwidth,angle=0]{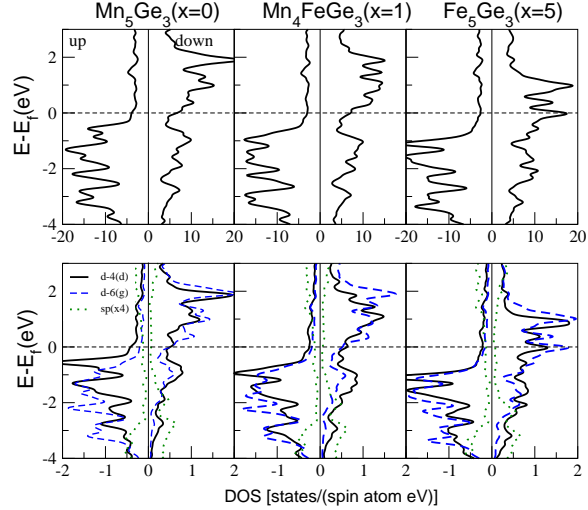}
\caption{(color online) Electronic density
 of states (DOS) of Mn$_{5}$Ge$_{3}$, Mn$_{4}$FeGe$_{3}$ and Fe$_{5}$Ge$_{3}$.
 Total and projected DOS onto metal $4(d)$ and $6(g)$ sites are shown in top and bottom panel.
 The $sp$-DOS has been multiplied by a factor of 4 for clarity.}
\label{fig:dos}
\end{figure}
  Fig.~\ref{fig:dos}  shows  the spin resolved total DOS (upper panels)
 and the DOS  projected onto $d$-states of 4(d) and 6(g) metal atoms as well as onto
  $sp$ states (lower panels). From left to right, the iron concentration varies from
   x=0,1 to x=5.  Let us focus on the region
  around the Fermi level for Mn$_{5}$Ge$_{3}$ (upper panel):
  while the spin up DOS has an almost structureless plateau
  (there is a small valley in the DOS curve just above the Fermi level),
the spin down DOS has much more fine structure  around and
immediately above the Fermi energy.  From the lower panel, we   see:
i) the $sp$ electrons are mainly located in the region below -2 eV
(bonding states) and above 1.5 eV (antibonding states), supporting
the picture given above; ii)  N(E$_{f}$) is   dominated by $d$
states rather than by $sp$ states, in both spin channels, suggesting
that the $d$  electrons  are mainly responsible for  the spin
polarization; iii) the number of $sp$-electrons with spin-up,
N$^{\uparrow}_{sp}$,
 is significantly larger than those with spin-down,
N$^{\downarrow}_{sp}$, suggesting that the average value  of the
spin-up Fermi velocity is  larger than the corresponding spin-down
value.\footnote{Remember that $sp$-electrons are expected to have
large velocity.} In the majority component, the $d$-4(d) DOS is
larger than the $d$-6(g) one, while  they are practically equal in
the minority component. Upon Fe doping (electron-like carrier
doping), the Fermi level shifts upwards to accomodate the extra
electron per formula unit and now lies at the bottom of the small
valley of the majority  DOS and on top of a small peak in the
minority DOS:
 as a result, N$^{\uparrow}$ decreases, and N$^{\downarrow}$ increases,
leading to a more negative spin-polarization P$_{0}$, which is
proportional to  N$^{\uparrow}$-N$^{\downarrow}$ (see below).
 These modifications
are mainly driven  by the $d$-4(d) states and are consistent with a
simple rigid band picture. On the other hand, there are negligible
changes in N$^{\uparrow}_{sp}$ and N$^{\downarrow}_{sp}$ at E$_{f}$.

In the high doping limit, we observe that the majority DOS is hardly
modified compared to the lower Fe content case, while the minority
component becomes more structured; in fact, the Fermi level is
further shifted upwards: it still lies at the bottom of a valley in
the  majority DOS, while it is pinned at a high peak  in the
minority component, leading to a very large value of the total DOS
at E$_f$. The large Fe content mainly affects the minority spin DOS
and strongly modifies the overall transport properties through large
modifications of the minority spin component. As before, the
$sp$-DOS at E$_{F}$ is marginally affected.

\begin{table}
 \caption{Spin polarizations for different $x=0,1,5$. See text for further details.}
  \label{tab2}
 \begin{ruledtabular}
\begin{tabular}{lccccccccc}
Compound         & x$_{0}$   & y$_{0}^{\parallel}$ & y$_{0}^{\perp}$
&P$_{0}$(\%) &P$_{1\parallel}$(\%) & P$_{1\perp}$(\%)     &
P$_{2\parallel}$(\%)
& P$_{2\perp}$(\%) \\
  Mn$_{5}$Ge$_{3}$ ($x=0$)  &  0.55  & 2.4 & 2.1  &-30  &14&7&52&41\\

 Mn$_{4}$FeGe$_{3}$($x=1$) &  0.28  & 3.7 & 2.6  & -55 &3&-16&59&30\\

 Fe$_{5}$Ge$_{3}$  ($x=5$) &  0.13  & 0.5 & 0.7  &-77  &-88&-83&-95&-87\\
\end{tabular}
\end{ruledtabular}
\end{table}
Let us now consider   the spin-polarization (SP) of the different
compounds.
A common definition for SP is:\cite{PaperMazin}
\begin{equation}\label{Mazin}
P_{n}=\frac{N_{\uparrow}(E_{F})v_{F\uparrow}^{n}-N_{\downarrow}(E_{F})v_{F\downarrow}^{n}}
           {N_{\uparrow}(E_{F})v_{F\uparrow}^{n}+N_{\downarrow}(E_{F})v_{F\downarrow}^{n}}
\end{equation}
 where N and $v_{F}$ are the density of states (DOS)  and average
Fermi velocity of electrons with spin-up, spin-down at $E_{f}$,
respectively. Note that Eq.~\ref{Mazin} can be written as follows:
 $P_{n}=(x_{0}y_{0}^{n}-1)/(x_{0}y_{0}^{n}+1)$
  with $x_{0}=N^{\uparrow}(E_{F})/N^{\downarrow}(E_{F})$ and
 $y_{0}=v_{F}^{\uparrow}/v_{F}^{\downarrow}$. Due to the hexagonal symmetry, the Fermi velocity can
 be decomposed into components which are  parallel and perpendicular to
the basal plane,
 $v_{F,\parallel}^{\uparrow,\downarrow}$  and $v_{F,\perp}^{\uparrow,\downarrow}$ respectively.
 The same is true for $y_{0}^{n}$.

 For $n=0$, the SP calculated using Eq.~\ref{Mazin} corresponds to spin-resolved
 photoemission measurements, while higher orders of $P_{n}$ correspond to SP
as measured in transport experiments, such as PCAR (point contact
Andreev reflection ) and TJ (tunnel junction), in the
 ballistic ($P_{1}$) or diffusive regime ($P_{2}$).\cite{SP}
 In Tab.~\ref{tab2}, we show the calculated $x_{0}$ and $y_{0}$ at $E_{f}$
  for
 the three compounds, and the corresponding  P$_{n}$ for $n$=0,1,2. Upon iron doping,
  we see that:
  i) N$^{\uparrow}<$N$^{\downarrow}$ ({\it i.e.} $x_0 < 1$) in all compounds considered and,
  in particular, N$^{\uparrow}\ll$N$^{\downarrow}$ for $x=5$.
 ii) For both the parallel and perpendicular component,
 $v_{F}^{\uparrow}>v_{F}^{\downarrow}$ ({\it i.e.} $y_0 > 1$),
 except for Fe$_{5}$Ge$_{3}$ where the opposite is true. The average velocity for spin up
 $v_{F}^{\uparrow}$ is larger than $v_{F}^{\downarrow}$, because the conductivity is
 dominated by the $sp$ electrons, and in all Mn$_{5-x}$Fe$_{x}$Ge compounds
 N$^{\uparrow}_{sp}>$N$^{\downarrow}_{sp}$.  Fe$_{5}$Ge$_{3}$ is special,
 since
 N$^{\downarrow}_{d}$ is exceptionally large (see Fig. 1).

 Clearly, point i) leads to  negative P$_{0}$, with the absolute values increasing
 with  the Fe content; {\em i.e.}  at the  Fermi-level, the spin up density is smaller
than the spin down density, and this behaviour is enhanced by Fe
doping.
 However  P$_{1}$ is greater than P$_{0}$, and P$_{2}$ is greater than P$_{1}$,
 except for Fe$_{5}$Ge$_{3}$. This is related to ii),
 {\em i.e.} the spin up velocity is exceeds   the spin down velocity  ($y_0 > 1$)
 over-compensating the lower spin-up versus spin-down density at Fermi-level.
 The exception is Fe$_{5}$Ge$_{3}$ where $y_0 <1$, and hence $P$ slightly decreases
 with $n$ ( $P_0\gtrsim P_1 \gtrsim P_2$).

In summary, we have studied the hexagonal phase (D8$_{8}$) of the
Mn$_{5-x}$Fe$_{x}$Ge intermetallic alloy, focusing on the role of
the Fe substitution on the structural, electronic and magnetic
properties of the compound. We  found that: i) Fe substitution
preferentially occurs at the $4(d)$ site of the D8$_{8}$ structure;
ii) substitution of one Mn atom/f.u. in Mn$_{5}$Ge$_{3}$ enhances
the spin polarization P$_{0}$ in agreement with experiment, but iii)
substitution also  decreases P$_{1}$ and P$_2$. The last observation
is  at variance with  the experimental results.\cite{mn4feg3} The
reason for this  disagreement deserves further studies that should
include disorder as well as interface effects, which are beyond the
purpose of the present study. On the other hand, our
\emph{ab-initio} calculations indeed confirm that is possible to
tune the degree of spin-polarization in the different transport
regimes upon Fe doping. Remarkably, an even larger spin polarization
than that of Mn$_{4}$FeGe$_{3}$  can be achieved by further
increasing the Fe content, thus making the
Mn$_{1-x}$Fe$_{x}$Ge$_{3}$ intermetallic alloy very promising for
future spintronic applications.

\label{concl}

\acknowledgments
This work was supported by the Austrian {\em Fonds zur F\"orderung
der wissenschaftlichen Forschung}; by CNR-INFM through Iniziativa
Trasversale Calcolo Parallelo and by
Consorzio Gran Sasso through a computing grant at Centro Calcolo dei Laboratori
Nazionali del Gran Sasso (INFN).

\newpage

\bibliography{biblio}

\newpage

\end{document}